\documentclass[12pt]{article}
\usepackage[english]{babel}
\usepackage[utf8x]{inputenc}
\usepackage[T1]{fontenc}

\usepackage[a4paper,top=3cm,bottom=2cm,left=3cm,right=3cm,marginparwidth=1.75cm]{geometry}

\usepackage{amsmath}
\usepackage{graphicx}
\usepackage[colorinlistoftodos]{todonotes}
\usepackage[colorlinks=true, allcolors=blue]{hyperref}
\usepackage{times}
\usepackage{cite}
\usepackage{epsfig}
\usepackage{amsmath}
\usepackage{amssymb}
\usepackage{authblk}
\usepackage{color}
\usepackage{soul}
\usepackage{epstopdf}
\usepackage{cancel}
\usepackage{hyperref}
\usepackage{url}

\def\mathswitch#1{\relax\ifmmode#1\else$#1$\fi}
\def\mathswitchr#1{\relax\ifmmode{{#1}}\else${#1}$\fi}
\newcommand{\PW}{\mathswitchr W}
\newcommand{\PZ}{\mathswitchr Z}

\newcommand{\Pt}{\mathswitchr t}

\newcommand{\Ps}{\mathswitchr s}
\newcommand{\MW}{\mathswitch {M_\PW}}

\newcommand{\GZ}{\mathswitch {\Gamma_\PZ}}

\newcommand{\mt}{\mathswitch {m_\Pt}}

\newcommand{\seff}[1]{\sin^2\theta_{\rm eff}^{#1}}

\newcommand{\as}{\mathswitch {\alpha_\Ps}}

\newcommand{\uncer}{{uncertainty}}

\newcommand{\mev}{\,\, \mathrm{MeV}}

\def\order#1{\ensuremath{{\cal O}(#1)}}

\definecolor{orange}{rgb}{0.8,0.5,0}

\title{{ \bf Theory Requirements and Possibilities for the FCC-ee and other Future High Energy and Precision Frontier Lepton Colliders\footnote{Input to the European Strategy Particle Physics 2018-2020}
}}

\author{Alain Blondel (Universit\'e de Gen\'eve), Ayres Freitas (University of Pittsburgh), 
 Janusz Gluza\footnote{Corresponding author: janusz.gluza@us.edu.pl}\; and Tord Riemann (U. Silesia),\vspace*{-.4cm}\\
 Sven Heinemeyer (IFT/IFCA CSIC Madrid/Santander, ECI/UAM/CSIC Madrid),\\ \vspace*{0cm} Stanis\l{}aw Jadach (IFJ PAN Krak\'ow), Patrick Janot (CERN)
 }

\date{\small 18 December 2018}

\begin{document}

\maketitle
\noindent

\begin{abstract}
The future lepton colliders proposed for the High Energy and Precision Frontier set stringent demands on theory. The most ambitious, broad-reaching and demanding project is the FCC-ee. We consider here the present status and requirements on precision calculations, possible ways forward and novel methods, to match the experimental accuracies expected at the FCC-ee. We conclude that the challenge can be tackled by a distributed collaborative effort in academic institutions around the world, provided sufficient support, which is estimated to about 500 man-years over the next 20 years. 
\end{abstract}
\vfill\eject
\section{Shaping the future of particle physics theory with FCC-ee}

The FCC project proposes to build a sequence of colliders in a common infrastructure close to CERN. It is a unique opportunity to step up in our understanding the physical world at the smallest scales. 
It is {\bf the most versatile}, 
feasible high energy collider project for the next few decades. 
Two, or even four, interaction points on the 100 km facility 
(important for setting up independent experimental  analyses) 
along with at least three different types of collisions ($ee$ as a first step, $hh$  and $eh$ in a second life, and may be more) will challenge the physics community with a very broad program.   
Fine details of the Standard Model theory 
of particle interactions will be revealed; even more importantly,
the widest possible spectrum of new theoretical concepts in the fundamental laws of physics will be explored. This quest will take several forms: i) exquisite precision measurements, ii) search for tiny violations of conservation laws, or iii) discovery of extremely rare phenomena forbidden in the Standard Model. Details can be found in \cite{FCC:cdrs2018}.

The following {\bf three characteristics} make 
the FCC multi-task and multi-purpose circular collider 
project special in its theoretical challenges. They call for a proper strategy for theoretical computations to back the analysis and interpretation of FCC-ee data.  

\begin{enumerate}
\item
{\bf FCC-ee: the first stage of the most powerful high energy discovery tool.}
The FCC-ee will provide a set of ground-breaking measurements  of a large number of new-physics sensitive observables, 
with  improvement by one to two orders of magnitude in experimental precision. {\bf This will require the corresponding improvement in calculation ability  
with respect to the present baseline of perturbative Quantum Field Theory (pQFT).}  
 An important feature of FCC-ee will be to improve the precision of input parameters of the SM, 
in particular $m_Z, m_{t}$ and $\alpha_{QCD}(m_Z)$; 
it will also offer a unique possibility to measure {\em directly} $\alpha_{QED}(m_Z)$,
without the use of low energy hadronic data.
{\bf Theoretical calculations will have to keep pace with these improvements.} {More precise knowledge of electroweak, Higgs 
and flavour physics will open the way to accurate investigations of the most burning questions in particle physics:  like the baryon asymmetry of the Universe, the nature of Dark Matter, the origin of neutrino masses, to name a few. See FCC CDR vol.~1~\cite{FCC:cdrs2018} for more extensive discussions.}  
\\      

 \item 
 {\bf{FCC-ee: an interdisciplinary challenge.}}
An interdisciplinary character of the FCC-ee project is manifest in the accelerator itself and the  experiments where cutting edge technologies,
equipment, facilities and software must be developed, often in the cooperation with the industry.\\
Presently available techniques of pQFT are unable to match 
the experimental precision of the future FCC data for almost all  observables.  The needs are well defined, especially in the case of the most demanding FCC-ee Tera-Z case \cite{Blondel:2018mad,mini}.
Future theoretical computations for FCC-ee physics will require 
{\bf a considerable synergy of pQFT with advanced and explorative mathematics}
in order to develop new computational methods to deal
with much higher level of complexity of the anticipated pQFT calculations.
Development of the new sophisticated  (Monte Carlo) programs for 
direct simulation of the scattering processes at FCC-ee will provide  another important
{\bf synergy with the corresponding branch of the computing science}. This connection between advanced pQFT and advanced exploratory mathematics
 and computing has played a fertile role in the development of both branches of science for a long time. 
New challenges in pQFT will induce 
significant progress in the computer algebra systems, algebraic geometry --
new cutting edge algorithms for analytical and numerical methods will be developed,
{\bf{to the profit of several branches in advanced mathematics and computing science.}}

\item 
	{\bf FCC-ee: highly demanding collective educational, scientific and social networking.} The enormous diversity of the physics topics that will be explored 
at several stages on the FCC constitutes a challenge 
both to theory and experiment parts of the grand project,
concerning a proper organization of the work and the choice of dedicated teams.
The project needs skillful and versatile researchers, 
in all stages of their scientific career.
Well prepared {\bf educational and research theoretical programs must be set up}, 
including Initial Training Networks, ERC and national grants. 
A world-wide collaborative  
effort of theoretical physicists and contributors from related
relevant disciplines will be carried out mainly in the framework of the academic environment. 
An initial  estimate of the efforts related to development of {\bf new theory
calculation techniques and tools for FCC experiments points to about 500 person years}, 
corresponding to about 25 full-time-equivalent TH experts throughout the design, 
construction and operation phases of the FCC-ee programme.
\end{enumerate}

\section{Precision theoretical issues and objectives at the FCC-ee}

FCC-ee defines a very clean experimental set-up compared to hadron colliders, but the demands in precision are much higher. Initial state QED radiation
is well understood, leading to the possibility of very precise measurements {\bf
requiring  leap-jumps in the precision of theoretical computations for Standard Model phenomena and so-called  
{\em electroweak pseudo-observables}} (EWPOs).   
Such are partial widths and couplings of Z and W, 
forward-backward and polarization asymmetries, peak cross sections at the Z resonance position,
effective electroweak mixing angle  {{~\cite{ALEPH:2005ab,Blondel:2018mad}}},
and additional EWPOs in the  WW, ZH and $t\bar{t}$ processes. 
EWPOs encapsulate experimental data after extraction of well known and controllable QED and QCD effects, 
in a model-independent manner.
They provide a convenient bridge between real data 
and the predictions of the SM (or SM plus New Physics).
Contrary to raw experimental data (like differential cross sections),
EWPOs are well suited for archiving and long term exploitation.
In particular archived EWPOs can be exploited over long periods of time
for comparisons with steadily improving theoretical calculations of the SM predictions,
and for validations of the New Physics models beyond the SM.
They are also useful for comparison and combination of results from different experiments.

For extracting the interesting physics from data in form of EWPOs
{\bf it will be mandatory to develop a library of new Monte Carlo (MC) event
generators} implementing both QED and EW/QCD higher order effects.
They will have to be developed for all processes 
at FCC-ee in all four stages, Tera-Z, WW, HZ and $t\bar{t}$.
Well tested legacy MC programs of the LEP era can be used for benchmarking purposes, but the 
development of new event generation techniques and programs, 
should be done in parallel with a campaign of advances in the theoretical physics  calculation methods.

Fitting EWPOs to data near Z-resonance was possible at LEP using 
non-MC semi-analytical calculations of QED effects 
{implemented in ZFITTER and TOPAZ0 programs}, and MC generators such as KORALZ.
However, already at LEP the extraction of W mass from WW process was
done by means of a direct fitting of the output of the MC programs (with W mass as a parameter) to WW cross section and final four fermion distributions, without any use of analytical formulas, with all QED effects handled by the MC {~\cite{Schael:2013ita}}. 
For FCC-ee data analysis, due to the rise of non-factorisable QED effects 
above the experimental errors, 
the latter way of using MC programs might become the standard for fitting EWPOs to data, even at the Tera-Z stage. 
New MC event generators will have to provide built-in provisions for
efficient direct fitting of EWPOs to data, which are not present in the LEP legacy MCs.
Section C3 of~\cite{Blondel:2018mad} describes 
possible forms of future EWPOs at FCC-ee experiments 
and specification of a new required MC software is given.
It is underlined there that
due to non-factorisable QED contributions, the factorization of the multiphoton QED effects 
will have to be done at the amplitude level. Additional quantities available in tau and heavy flavour physics will reach the $10^{-5}$ precision and are likely to need similar attention.

The state of art and the open issues for these analysis in the context of FCC-ee near $Z$-boson resonance 
has been reported in  \cite{Blondel:2018mad}. 
As FCC will start from $e^+e^-$ Tera-Z option
and it will bring the highest luminosities and stringiest experimental results, 
{\bf we quote below the main goals and tasks}, 
adapted from the executive summary given in \cite{Blondel:2018mad}.
\begin{enumerate}
\item 
In order to meet the experimental precision of the FCC-ee Tera-Z for EWPOs, 
3-loop and partial 4-loop calculations of the $Zf{\bar f}$-vertex will be
needed. The leading 3-loop corrections are
${\cal{O}}(\alpha \alpha_s^2), 
{\cal{O}}(N_f \alpha_{}^2 \alpha_s),
{\cal{O}}(N_f^2 \alpha_{}^3)$,
where $\alpha$ denotes an electroweak loop, and $N_f^n$ denotes $n$ or more
closed internal fermion loops. Sub-leading corrections may also be needed,
depending on insights gained from computing the leading terms.
\item  
Full 2-loop corrections to the $Zf{\bar f}$-vertex have been completed recently.
The principal techniques for the electroweak loop calculations have been and
will likely be numerical, 
due to the large number of scales involved.
A few digits of internal precision will be sufficient at the 3-loop level.
Several additional exploratory strategies, methods and tools 
have been also identified and discussed.
\item
The $Zf{\bar f}$-vertex corrections are embedded in a structure describing 
the hard scattering process $e^+e^- \to f  {\bar f}$, 
based on matrix elements in form of Laurent series around the $Z$ pole. 
Here, additional non-trivial contributions like 2-loop weak box diagrams must be implemented properly. 
\item
The experience gained in the high precision calculation for physics near Z pole  will provide a strong basis for
more advanced treatment of four fermion processes
in the  W mass and width measurements.
In spite of the fact that the experimental precision is less demanding, (permil precision rather than a few per-million), more
precise SM calculations will be required for the Higgs and top physics, as well. 
\item 
Numerically dominant effects due
multi-photon emission constitute another key problem;
their complexity will be often comparable 
to that of the electroweak loop calculations.
The methodology of joint treatment of electroweak and QCD loop corrections 
with the photonic corrections is essentially at hand. 
In practice, however, construction of new Monte Carlo programs, which
can handle efficiently the above problems, will need special efforts \cite{Jadach:2018lwm}.
\end{enumerate}

The elementary techniques for higher order perturbative
SM corrections are basically understood, 
but their use in practical calculations at the level of computer
programs will be highly nontrivial and will require considerable effort.
The understanding of all sources of possible theoretical uncertainties will be fundamental for success of the FCC-ee data analysis.

\begin{table}[h!]
\begin{center}
\begin{tabular}{|c|c|c|c|c|c|}
\hline
& $\delta \Gamma_Z\;[\rm MeV]$ &  $\delta R_l \; [10^{-4}]$ & $\delta R_b\; [10^{-5}]$ & 
			$\delta \sin^{2,l}_{eff} \theta \; [10^{-6}]$
\\ \hline
\multicolumn{5}{|c|}{Present EWPO theoretical  uncertainties} 
\\
\hline
EXP-2018 & 2.3 & $250$ & $66$ & $160$ 
\\
TH-2018   & 0.4  &  60   & $10$ & $45$
\\  
\hline 
\multicolumn{5}{|c|}{EWPO theoretical  uncertainties when FCC-ee will start}  
\\
\hline
EXP-FCC-ee   & 0.1 & 10 &  $2\div 6$  & $6$
\\
TH-FCC-ee   & $0.07$ & $7$ & $3$ & $7$ \\
\hline
\end{tabular}
\end{center}
\vspace{-2ex}
\caption{\it Comparison for selected precision observables 
of present experimental measurements (EXP-2018),
current theory   errors (TH-2018), FCC-ee precision goals at the end of the Tera-Z run (EXP-FCC-ee) 
and rough estimates of the theory  errors assuming that 
electroweak 3-loop corrections  and  the 
dominant 4-loop EW-QCD corrections are available at the start of FCC-ee (TH-FCC-ee).
Based on discussion in~\cite{Blondel:2018mad}. 
\label{tab:errors}}
\end{table}

Table~\ref{tab:errors} shows
the current experimental and theoretical errors (EXP-2018, TH-2018)  for some basic Z-physics EWPOs,
and the prospective measurement errors at FCC-ee (EXP-FCC-ee) 
together with the corresponding
estimate for theoretical uncertainties after the leading 3/4-loop results
become available (TH-FCC-ee). 
The entry TH-2018 takes into account recent completion of the 2-loop electroweak 
calculations~\cite{Dubovyk:2016aqv,Dubovyk:2018rlg}, 
so the error estimate comes solely from 
an estimate of magnitudes of missing 3-loop and 4-loop EW and mixed EW-QCD corrections. 
The estimated TH-FCC-ee error stems from 
remaining 4-loop and higher effects. 
These are rather difficult to estimate presently, 
however, a rough conservative upper bound on them has been provided in \cite{Blondel:2018mad}. 
They are denoted in Tab.~\ref{tab:errors} as TH-FCC-ee. 
As we can see, {\bf the uncertainties of the
TH-FCC-ee scenario are comparable to the corresponding EXP-FCC-ee experimental
errors and will not limit the FCC-ee physics goals}. 
\begin{table}[h!]
\begin{center}
\begin{tabular}[c]{|c|c|c|c|}
\hline
\multicolumn{4}{|c|}{{\large{ $Z \rightarrow b \bar{b}$}}}
\\
\hline
& 1 loop  & \hspace{0.9cm}2 loops\hspace{0.9cm} &  \hspace{1.2cm}3 loops \hspace{1.2cm} \\
Number of topologies
& 1 & 5   & 51\\
\hline
{Number of diagrams}& 15 & 1074 & 120472\\
\hline
{Fermionic loops} &0 & ${150}$ & $17580$\\
\hline
{Bosonic loops} &15 & ${924}$ & ${102892}$\\
\hline
{QCD / EW} &1 / 14&{98 / 1016}& ${10386 / 110086}$\\
\cline{2-4}
\hline
\end{tabular}
\end{center}
\caption{\it Numbers of relevant topologies and diagrams  for $Z\to b \bar{b}$
decays.
Adapted from~\cite{Blondel:2018mad}.} 
\label{tab:diagrams}
\end{table}

An illustration of complexity of future perturbative calculations
is provided in Table~\ref{tab:diagrams}, where
the numbers of distinct topologies  of diagrams to be calculated and
the numbers of diagrams and various categories
at the 1-, 2- and 3-loop order are shown for the most complicated $Z \to b \bar b$ decay.
According to Ref.~\cite{Blondel:2018mad}, 
the calculation of fermionic 3-loop corrections, 
which will be the largest part of them,
are within reach, already with present methods and tools.

Another important issue in electroweak fits are uncertainties from the existing (often experimental) errors on SM input
parameters, themselves EWPOs, which affect the SM predictions of EWPOs when comparing them with experiment. 

Well-defined subsets of experimental data (EWPOs) 
are treated as input SM parameters, 
like~\cite{deBlas:2016ojx,Tanabashi:2018oca} for instance: 
Fermi constant (from muon decay), $\alpha_{QED}(M_Z)$ (from the dispersion integral of $e^+e^-$ data at low energies), $\alpha_s(M_Z)$ (from the measurement of $R_\ell$ and $R_b$ at the Z peak),
$M_Z$, $m_t$, $M_H$ and the masses of other fermions --
all of them with well defined experimental and theoretical errors.
Remaining EWPOs, like $\sin^2\theta^{eff}_{W}$, $m_W$,
$\Gamma_Z$, and more, 
can be  measured either {\em directly} with an experimental error 
or can be predicted  using SM perturbative machinery,
with {\em parametric errors}  inherited from the chosen subset of input parameters (EWPOs).

\begin{table}[ht!]
\centering
\renewcommand{\arraystretch}{1.1}
\begin{tabular}[t]{l|cccc}
\hline
EWPO & Exp. direct error & Param. error & Main source & Theory uncert. \\
\hline
$\GZ$ [MeV] & 0.1 & 0.1             & $\delta\as$ & 0.07 \\
$R_b$ [$10^{-5}$] & 6 & $1$         & $\delta\as$ & 3 \\
$R_\ell$ [$10^{-3}$] & 1 & $1.3$    & $\delta\as$ & 0.7\\
$\seff{\ell}$ [$10^{-5}$] & 0.5 & 1 & $\delta(\Delta\alpha)$ & 0.7 \\ 
$\MW$ [MeV] & $0.5$ & 0.6 
& $\delta(\Delta\alpha)$ & 0.3  \\ 
\hline
\end{tabular}
\caption{Estimated experimental precision for the direct 
measurement of several important EWPOs at FCC-ee~\cite{Blondel:2018mad}
(column two) and  experimental parametric error (column three),
with the main source shown in the forth column.
Important input parameter errors are
$\delta(\Delta\alpha)=3 \cdot 10^{-5}$, $\delta\as=0.00015$
see FCC CDR, vol.~2~\cite{FCC:cdrs2018}.
Last column shows anticipated theory uncertainties at start of FCC-ee.  
}
\label{tab:paraerr}
\end{table}

In Table.~\ref{tab:paraerr} the size and main sources of parametric errors connected with QED and QCD running couplings for selected important EWPOs are shown. It is important to note that e.g. $\delta\as$ is a common parametric error on $\GZ$ and $R_\ell$, and will not affect e.g. the leptonic partial width of the Z. These EWPOs serving as SM input parameters
will be determined very precisely at the FCC-ee. 
It should be noted that determination of $\Delta \alpha$ 
with error at the level of $3\cdot 10^{-5}$ assumed in the table
is from present perspective a big challenge.
Standard approach exploiting low energy hadronic contributions to vacuum polarization
will try to reach such a precision.
Another independent method aiming at the same precision was proposed
\cite{Janot:2015gjr}.
It relies on measuring at FCC-ee charge asymmetry of the $e^+e^- \to \mu^+ \mu^-$ process
very precisely at $s^{1/2} \simeq M_Z \pm 3.5$~GeV.
Future projections of the theory uncertainty are also shown in Table.~\ref{tab:paraerr}. 
{\bf Both theoretical uncertainties and parametric errors  are smaller or comparable 
to the corresponding experimental direct errors.}

To complete the picture of precision studies at the FCC-ee, 
the remaining FCC-ee-W, FCC-ee-H and FCC-ee-t stages of  
the $e^+e^-$ collider should be also considered.

Very precise determinations of $M_W$
at FCC-ee will rely on the precise measurement
of the cross section of the $e^+e^- \to W^+W^-$ process near the threshold.
A statistical precision  {of} $0.04\%$ of this cross section translates
into 0.6~MeV experimental error on $M_W$,
which is almost three orders of magnitude better than at LEP.
{Theoretical legacy calculations from the LEP era} for this process with
a precision tag of $\sim 0.2\%$ would
yield an unacceptable 3 MeV theory error  for  $M_W$ \cite{Actis:2008rb}.
Therefore, improved theoretical calculations are required 
for the generic $e^+e^- \to 4f$ process near the WW threshold with an improvement of one order of magnitude.  As advocated in~\cite{Skrzypek:Jan2018},
the most economical solution will be to combine \order{\alpha^1}
calculation for $e^+e^- \to 4f$ process with \order{\alpha^2}
calculation for the doubly-resonant $e^+e^- \to W^+W^-$ subprocess.
The former calculation is already available~\cite{Denner:2005fg}.
The latter will need to be developed; Inclusion of the resummed QED corrections will be mandatory.

{\color{black}{Present estimation for the above two-loop electroweak \order{\alpha^2} corrections based on the unstable-particle Effective Field Theory approach  translates into 1 MeV theory uncertainty of $M_W$ \cite{Beneke:2007zg,Schwinn2013}.}}
Assuming that the two-loop electroweak \order{\alpha^2} corrections are known, an estimate of the missing three-loop electroweak \order{\alpha^3} corrections 
in such a hybrid calculation~\cite{Skrzypek:Jan2018},
shows that $\sim 0.02\%$ theory precision for the threshold cross section
could be achieved, which translates into  0.3 MeV  theory uncertainty of $M_W$ given in Tab.~\ref{tab:paraerr}, 
comparable with the FCC-ee experimental precision.
On the other hand, $M_W$ mass will be also extracted
from the subsequent decays of W's  to four fermions, away from the threshold.
Most likely, the same progress in the theory calculations
will be mandatory for this measurement as well.

In case of FCC-ee-H, $M_H$ will be obtained from $e^+e^- \to HZ$ process
with the precision better of order 10 MeV \cite{FCC:cdrs2018}. 
Theory uncertainties 
(mainly due to final-state radiation effects) will be subdominant. 

The anticipated experimental error
for $m_t$ measurement at FCC-ee-t~\cite{Blondel:2018mad}
is $\delta \mt \sim \order{20}\mev$.
On the theory side, there are several \uncer\ sources: 
\emph{(i)} 
The perturbative uncertainty
for the calculation of the threshold shape with higher order QCD corrections,  
\emph{(ii)} 
the threshold mass definition translated into the
$\overline{\text{MS}}$ scheme
\emph{(iii)} precision of $\as$.  
Combining these three error sources,
a theory uncertainty close to experimental and less than $50$~MeV for $\mt$  appears feasible. 
In addition a very accurate determination of the efficiency 
of experimental acceptances and selection cuts is needed. 
This task will require the inclusion of higher-order corrections 
and re-summation results in a Monte-Carlo event generator. 
NLO QCD corrections for off-shell $t\bar{t}$ production, and matching between these contributions  are complementing previous semi-analytic results. 
 
An initial workshop "Precision EW and QCD calculations for the FCC studies: 
methods and tools" was held at CERN in January 2018 
\cite{mini} to start the process of preparing long-term research for precision calculations needed in the FCC-ee Tera-Z physics. This activity will continue in 2019 \cite{mini2019}; this follow-up workshop  will focus  on establishing solid plans for precision calculations the FCC-ee project connected with W, H and top quark physics. 

\section{The need for development of methods and 
refined tools in theoretical calculations} \label{mtools}

An estimation of theory errors has always to be taken with a grain of salt.
As was shown recently in \cite{Dubovyk:2018rlg} they can differ from the actual calculations even by factors 3-5.
There is no other way than to perform the concrete calculation of higher loop effects, and, in view of the huge complexity, ideally with at least two independent methods and/or teams.
In fact, we will need the leading electroweak 3-loop and QCD 4-loop contributions in order not to limit the interpretation of the Z resonance shape, and there are similar demands from the other FCC-ee modes.  
There is no closed form theory for perturbation theory calculations of Feynman integrals beyond one loop.
For this reason, numerical integration methods are the most promising, if not the only, avenues for addressing that challenges.
Analytical techniques are expected to be important in many respects, but
numerical integration methods have advantages 
for increasing number of masses and momentum scales.
Fortunately, there is an impressive progress in recent years in this direction \cite{Blondel:2018mad}.
The renormalization and treatment of ultra-violet divergences is basically understood.
There are only two numerical methods known to allow a systematic treatment of infra-red divergences. 
In 2014  the only advanced  automatic numerical 
two-loop method was one of them, sector decomposition  {SD}.
However, the corresponding software was not capable to work out 
the complete set of Feynman integrals for the electro-weak bosonic two-loop corrections 
to the $Z$-boson decay with the desired high precision (up to 8 digits per integral). 
The task could be completed succesfully with a substantial development of the
other numerical approach, based on  Mellin-Barnes  {MB} representations of Feynman
integrals~\cite{Dubovyk:2016aqv,Dubovyk:2018rlg}. 
These calculations are challenging due to the numerical role of particle masses $M_Z, M_W, m_t, M_H$, leading to
(i)
a huge number of contributions, ranging from tens to hundreds of thousands of diagrams (at 3-loops),
and (ii)
the occurrence of up to four dimensionless parameters in Minkowskian kinematics (at $s=M_Z^2$)  with intricate threshold and on-shell effects where contour deformation fails. 
To tackle more loops or legs, merging both the MB- and SD-methods in numerical calculations is a key for final success.

A well founded expectation is that these two methods, combined with more
specialized techniques for specific topologies, will allow to execute the next steps of the necessary 
multi-loop calculations.
There are  also many opportunities for improvement in other approaches and software  packages; 
see \cite{Blondel:2018mad} for a year-2018 overview. 
There is a lively activity in 
many areas of multi-loop calculations,
and several promising (semi-) analytical methods 
are also under development \cite{Blondel:2018mad}. 
For instance a treatment of multi-scale integrals beyond multiple polylogarithms,
direct numerical calculations of Feynman integrals in $d=4$,
or calculations based on unitarity methods, 
all of them with first examples of two-loop numerical evaluations. 

Concerning the numerically important QED corrections, further refinements of
the factorization to infinite order of the multi-photon soft real and virtual contributions, and their re-summation, using Monte Carlo methods have to be pursued in order to meet the accuracy needs.
Again infra-red problems appear to be crucial: The inclusion of collinear (mass) singularities must be improved
in the existing schemes.
Techniques for disentangling QED and EW corrections for two and more loops in the
framework of soft photon factorization/resummation are in principle available,
but their practical implementation will require more work.
A central issue is the implementation of the higher loop terms in their proper
form as Laurent series around the Z-boson pole, characterized by mass and width, 
in the scattering energy, which is mandated by analyticity, gauge-invariance and unitarity. 
 
\section{Summary}
FCC-ee, a circular collider with extremely high statistics and high energy
resolution, will provide the possibility to test the Standard Model with its fine quantum electroweak effects 
{\bf with a precision far beyond the current state of the art.} 
Significant future theory effort will be needed for
both for parametric and theoretical calculational errors to match the experimental accuracy of FCC-ee physics program. {\bf No potential showstoppers are foreseen} \cite{FCC:cdrs2018,Blondel:2018mad}. It will be important that adequate theory funding will be available
to ensure that theory uncertainties are reduced to the desired level.
The combined advances in experimental and theoretical
techniques will yield unprecedented sensitivity to very weakly coupled or very massive new physics, or to forbidden violations of symmetries, at FCC-ee.

\section{Acknowledgments}
The authors acknowledge the contributions of all participants in the FCC-ee working groups, especially the participants of the workshop on theory calculations~\cite{mini}.

{\textit{A.B.} gratefully acknowledges CERN hospitality. The work of \textit{T.R.}\ was supported in part by an Alexander von Humboldt Polish Honorary Research Fellowship.
	The work of   \textit{J.G.} was supported by the Polish National Science Centre (NCN) under the Grant Agreement 2017/25/B/ST2/01987.
 The work of \textit{A.F.}\  was supported in part by the National Science Foundation under
 grant no.\ PHY-1519175. The work of \textit{S.H.} was supported in part by the
MEINCOP (Spain) under contract FPA2016-78022-P, in part by the Spanish
Agencia Estatal de Investigaci\'on (AEI), in part by the EU Fondo
Europeo de Desarrollo Regional (FEDER) through the project
FPA2016-78645-P, in part by the ``Spanish Red Consolider MultiDark''
FPA2017-90566-REDC, and in part by the AEI through the grant IFT
Centro de Excelencia Severo Ochoa SEV-2016-0597.
 The work of \textit{S.J.} was partly supported by
 the Polish National Science Center grant 2016/23/B/ST2/03927.
 


\end{document}